\documentclass[aps,prl,twocolumn,showpacs,a4paper,superscriptaddress]{revtex4}
\usepackage{graphicx}
\usepackage{amsmath}

\begin{document}
\title{Ab initio GW electron-electron interaction effects in Quantum Transport}

\author{Pierre \surname{Darancet}} 
\affiliation{LEPES, UPR 11 CNRS, 38042 Grenoble, France}
\affiliation{European Theoretical Spectroscopy Facility (ETSF)}
\author{Andrea \surname{Ferretti}} 
\affiliation{Dipartimento di Fisica, Universit\`a di
             Modena e Reggio Emilia, and
             INFM-CNR-S3, National Center on nanoStructures
             and bioSystems at Surfaces, 41100 Modena, Italy}
\affiliation{European Theoretical Spectroscopy Facility (ETSF)}
\author{Didier \surname{Mayou}} 
\author{Valerio \surname{Olevano}} 
\affiliation{LEPES, UPR 11 CNRS, 38042 Grenoble, France}
\affiliation{European Theoretical Spectroscopy Facility (ETSF)}
\date{\today}

\begin{abstract}
We present an {\it ab initio} approach to electronic transport in
nanoscale systems which includes electronic correlations
through the GW approximation. 
With respect to Landauer approaches based on density-functional theory (DFT),
we introduce a physical quasiparticle electronic-structure into 
a non-equilibrium Green's function theory framework.
We use an equilibrium non-selfconsistent $G^0W^0$ self-energy
considering both full non-hermiticity and dynamical effects.
The method is applied to a real system, a gold mono-atomic chain.
With respect to DFT results, the conductance profile is modified 
and reduced by to the introduction of diffusion and loss-of-coherence
effects. The linear response conductance characteristic
appear to be in agreement with experimental results.

\end{abstract}

\maketitle

Electronics at the nanoscale, namely {\it nanoelectronics},
represents the next years' technological challenge.
It is boosted not only by the need for shorter integration scales,
but also by the expectation that unusual quantum effects ~\cite{QTnatures} 
are going to be observed due to quantum phenomena effects. 
Beside the experimental efforts to synthesize nanoelectronic devices,
quantum transport theory~\cite{Datta} has the formidable task to understand and to model the
mechanisms behind these phenomena and to predict them from
a first principles approach.

In the last years, a combination of \textit{ab initio} 
density-functional theory (DFT) calculations
together with the description of transport
properties in a Landauer-B\"uttiker (LB) framework~\cite{Datta} 
has demonstrated its ability to describe
small bias coherent transport in nanojunctions~\cite{want,transiesta,nanotubes}.
These approaches were successful
in accounting for the contact resistance
and conductance degrading mechanisms induced by impurities, defects
and non-commensurability patterns in the conductor region.
The major objections raised to such method are:
($i$) the Kohn-Sham (KS) electronic structure is in principle unphysical, 
to be considered only as an approximation to the quasiparticle (QP)
electronic structure;
($ii$) non-coherent and dissipative effects due to
electron-phonon (\textit{e-ph}) and electron-electron (\textit{e-e}) scattering
can be taken into account only approximatively in the LB formalism;
($iii$) non-linear response and far from equilibrium finite-bias transport
are not accessible, since DFT cannot be applied to open systems
and is not a non-equilibrium theory
(although recent works~\cite{Kurth2005} have demonstrated
that {\it time-dependent} DFT can tackle the problem).

Non-equilibrium Green's function (NEGF) theory~\cite{KadanoffBaym,HaugJauho}
is in principle a correct approach to address the above objections.
The critical point within this theory is the choice of good approximations
to the self-energy $\Sigma^r$, and coherently to the scattering functions
$\Sigma^{<,>}$.
This ensures that both the renormalization of the QP energies and the electron
diffusion mechanisms due {\it e.g.} to \textit{e-ph} or \textit{e-e} interactions
will be properly taken into account.
First works studying the role of the 
\textit{e-ph} coupling~\cite{FrederiksenBrandbyge2004,Guo}
and of short-range \textit{e-e} correlations~\cite{DelaneyGreer2004,FerrettiCalzolari2005} 
or the renormalization of the QP energies~\cite{Pecchia}
have recently appeared in the literature, or, we are aware, are going to appear~\cite{Thygesen}.
The role of correlations, apart from being central in explaining 
{\it e.g.} Coulomb blockade and Kondo effects, 
could also be crucial in bridging the gap between experimentally-observed and
LB-predicted conductances, in some cases are orders of magnitude 
off~\cite{DelaneyGreer2004,Evers2004,Kurth2005}.

%
%
In this work, we introduce electronic correlations
in the calculation of transport by an {\it ab initio} approach
based on Hedin's
GW approximation (GWA)~\cite{Hedin1965,GW} in the framework of NEGF.
In our scheme, the GW self-energy is built at equilibrium and 
the Green's function is calculated by direct solution of the Dyson equation.
For the lead/conductor/lead geometry, the GW self-energy is summed
to the lead's self-energies; the electronic conductance is 
calculated through the Meir-Wingreen formula~\cite{MeirWingreen1992}, 
a NEGF Landauer-like expression derived for interacting conductors.
We apply this scheme to a realistic system, a gold mono-atomic unidimensional 
chain~\cite{AgraitYeyati2003}, and 
we study the effects induced on transport properties by the 
different components of the GW self-energy, 
the hermitean and the non-hermitean parts and the dynamical dependence. 
Our results show that the conductance profile
is considerably modified by the real part of the GW correction.
The imaginary part introduces a suppression of the conductance, which is negligible
close to the Fermi energy, but that increases with the energy. 
Finally, the full dynamical dependence of the GW self-energy introduces
further structures far from the Fermi energy,
which have to be ascribed to satellite excitations of the system. 
The GW smooth drop on the conductance characteristics 
as a function of the bias at very low voltage
compares favorably with the trend experimentally measured 
in gold nanowires~\cite{AgraitYeyati2003}.

\begin{figure}
  \includegraphics[clip,width=0.45\textwidth]{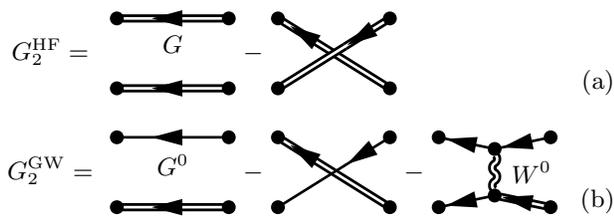}
  \caption{2-particle Green's function in the
  self-consistent Hartree-Fock (a) 
  and in the non self-consistent $G^0W^0$ approximation (b).
  Thin line: non self-consistent $G^0$; 
  double line: self-consistent $G$; 
  wiggly line: RPA non self-consistent $W^0$.
  }
  \label{g2}
\end{figure}

%
%
With respect to Hartree-Fock (HF), which already renormalizes
the energies for the \textit{e-e} classical repulsion and exchange, 
the GWA introduces the important contribution due to correlations
and to \textit{e-e} scattering diffusion mechanisms responsible for
loss-of-coherence in transport. 
Indeed, by direct inspection of the diagrammatic representation of
the 2-particle Green's function $G_2$
(see Fig.~\ref{g2} and Refs.~\cite{KadanoffBaym,SchindlmayrGarcia-GonzalezGodby2001}),
one can see that the $G_2^{\rm HF}$ describes an uncorrelated propagation and
that collisional terms are missing. 
This implies that the HF scattering functions, $\Sigma^<$ and $\Sigma^>$, 
are exactly zero. 
On the other hand, even a non-selfconsistent $G^0W^0$ approximation introduces
a collisional term [the last diagram in Fig.~\ref{g2}(b)] which gives rise to
non-zero $\Sigma^{<,>}$, and in turn to \textit{e-e} scattering
mechanisms and incoherent, dissipative effects in transport.
As it is shown by the corresponding $G_2$ Feynman diagram, the $G^0W^0$ approximation
is not a conserving approximation in the Baym and Kadanoff sense~\cite{KadanoffBaym},
leading to, e.g., non conservation of the number of particles. 
However, the relative deviation from the exact density brought by the 
$G^0W^0$ approximation has been evaluated by
Schindlmayr et al.~\cite{SchindlmayrGarcia-GonzalezGodby2001} to be only
of the order of 0.05\% for the range of metallic densities ($r_s^{\rm Au} = 3.01$)
of interest here.

Our starting point is a standard DFT-LDA calculation based on plane waves (PWs) and
norm-conserving pseudopotentials
for an infinite mono-atomic chain of gold atoms
using periodic boundary conditions~\cite{dftdetails}.
The KS electronic structure is calculated
both at the relaxed atomic distance (4.72 Bohr) and in a {\it stretched}
geometry (5.32 Bohr), so as to simulate the experimental
situation described in Ref.~\cite{AgraitYeyati2003} 
(conductance measures of a gold monoatomic chain pulled up
from a gold surface by an STM tip) and also the
calculations reported in Ref.~\cite{FrederiksenBrandbyge2004}.
>From the DFT KS eigenfunctions, we obtain an orthonormal
set of {\it maximally localized}
Wannier functions (MLWF)~\cite{MLWF},
which are used as a basis set in the calculation of quantum transport.
The following step is a converged~\cite{gwdetails} GW plane-wave calculation of
both the QP energies and the self-energy matrix elements for
the six bands around the Fermi energy, corresponding to the gold $sd$-manifold.
The self-energy in the $G^0W^0$ approximation at equilibrium is given by:
\begin{equation}
  \label{eq:GW}
  \Sigma_{\rm GW}(\omega) = \frac{i}{2 \pi} 
  \int_{-\infty}^{\infty} d\omega' \, 
  e^{-i \omega' 0^+} 
  G^0(\omega-\omega') 
  W^0(\omega')
  ,
\end{equation}
where $G^0$ is the Green's function built on the non-interacting
KS electronic structure and 
$W^{0}$  is the dynamically screened interaction given by the RPA polarizability,
$P^{\rm RPA}= -i G^0 G^0$.
Since for transport we need a fine-grid fully dynamical dependence
of the self-energy,
we calculate the frequency integral of Eq.~(\ref{eq:GW}) in three different ways:
({\it i}) by approximating the dynamical dependence
of $W(\omega')$ through a {\it plasmon-pole} 
(PP) model \cite{GW};
({\it ii}) by a {\it contour deformation} (CD) method \cite{cd}, 
which consists in a deformation of the 
real axis contour such that the self-energy can be calculated as an
integral along the imaginary axis
minus a contribution arising from the residual of the contour-included poles of $G$;
({\it iii}) by an {\it analytic continuation} (AC) method \cite{ac}, 
\textit{i.e.} calculating the integral and also the self-energy on the imaginary axis 
and then performing an analytical continuation to the real axis.
In the last step we carried out the quantum transport calculation
using a modified version of the \textsc{WanT} code~\cite{want,FerrettiCalzolari2005}.
We first projected the GW self-energy,
as well as the non-interacting hamiltonian $H^0$, on the Wannier functions basis set
(non-diagonal self-energy elements in the Bloch basis were neglected).
We study the bulk conductance and also
partition the system into three regions: the right (R) and left (L)
leads -- two semi-infinite gold mono-atomic chains -- and a central (C) region,
constituted by a single gold atom.
This has the purpose of clarifying the role of both intra-conductor
and conductor-lead correlations.
We calculate the retarded Green's function in the space spanned by the MLWF set by
inverting the Dyson equation, {\it i.e.}
\begin{equation}
  G^r(\omega) = [\omega - H^0 - \Sigma^r(\omega)]^{-1}
\end{equation}
where $H^0$ is the KS hamiltonian once
the exchange-correlation contribution is subtracted, $H^0 = H_{\rm KS} - V_{xc}$.
For the tri-partitioned geometry, the total retarded self-energy,
\begin{equation}
  \Sigma^r = \Sigma^r_L + \Sigma^r_R + \Sigma^r_{\rm GW}
  ,
\end{equation}
is the sum of the correlation GW and lead self-energies.
The conductance is finally calculated using the following formula:
\begin{equation}
  \label{eq:generalized_landauer}
  C(\epsilon) = \frac{2e^2}{h} 
    {\rm tr}[G^a \Gamma_R G^r \Gamma_L (\Gamma_L + \Gamma_R)^{-1} \Gamma]
    ,
\end{equation}
first given by Meir and Wingreen~\cite{MeirWingreen1992}
and recently re-derived under more general
conditions~\cite{FerrettiCalzolari2005}.
Here $\Gamma=\Gamma_L + \Gamma_R + \Gamma_{\rm GW}$, 
not to be confused with the vertex function,
is the {\it total decay rate} (the sum of the electron in- and out-scattering
functions), due to both the presence of the R and L leads and the effect of the
\textit{e-e} interaction,
\begin{equation}
  \Gamma = i (\Sigma^r - \Sigma^a) = i (\Sigma^> - \Sigma^<)
  ,
\end{equation}
With respect to the Landauer formula, Eq.~(\ref{eq:generalized_landauer}) 
presents a factor $(\Gamma_L + \Gamma_R)^{-1} \Gamma$
which reduces to one in the uncorrelated case.
Equation~(\ref{eq:generalized_landauer}) represents an {\it effective} 
(including correlation effects) transmission
probability of an electron injected at energy $\epsilon$ through the conductor.

\begin{figure}
  \includegraphics[clip, width=0.45\textwidth]{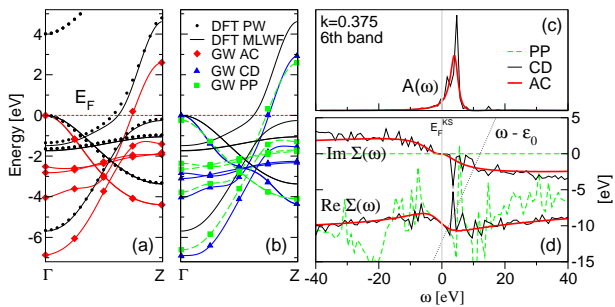}
  \caption{ (color online)
  (a) and (b): DFT-LDA Kohn-Sham vs GW electronic structure:
  black dots: DFT-LDA on PW basis;
  lines: DFT-LDA on MLWF;
  squares, triangles and diamonds refer to GW calculations from the PP model,
  the CD, and the AC methods respectively.
  Fermi energy is set to zero.
  (c) Spectral function and (d) 
  real and imaginary part of the GW self-energy:
  dashed, thin and thick lines are from the PP model, 
  the CD, and the AC methods respectively. 
  The dotted curve is the straight line $\omega - \epsilon^KS + \langle V_{xc} \rangle$
  whose intersections with the real part of the self-energy give the peaks of the
  spectral function.  
  The 0 is set to the Kohn-Sham Fermi energy.
  }
  \label{bandplotsigma}
\end{figure} 
In Fig.~\ref{bandplotsigma}(a,b) we compare the DFT-LDA (Kohn-Sham)
and the GW (quasiparticle) electronic structure for the relaxed geometry.
The dots represent the DFT-LDA levels from the PW calculation.
We verified that the diagonalization on the MLWF basis (solid lines) 
closely reproduces the PW results.
Squares, triangles, and diamonds
represent the GW electronic structures calculated using
the PP model, the CD, and the AC methods, respectively.
Little difference among the GW methods is found.
The GW corrections globally lower the $d$-like states wrt the Fermi level,
and also reduce the $s$-like bandwidth (this effect is less evident in
the stretched chain where the bandwidth shrinks about 3 eV
already at the DFT level).

In Fig.~\ref{bandplotsigma},
we compare the real and imaginary parts of the self-energy (d) and
the spectral function $A = i (G^r - G^a)$ (c) for a point close 
to the Fermi level. 
We remark that the AC method appears to smooth the 
richer-in-structure CD spectrum. Although computationally cheaper,
the AC method
should be considered less accurate than CD, especially on the imaginary part.
However, the frequency dependence as well as the shape and the position 
of the main structures (both QP and satellites peaks) 
are essentially caught by both methods.
Therefore, we use the AC approach in the following.

\begin{figure}
  \includegraphics[clip, width=0.45\textwidth]{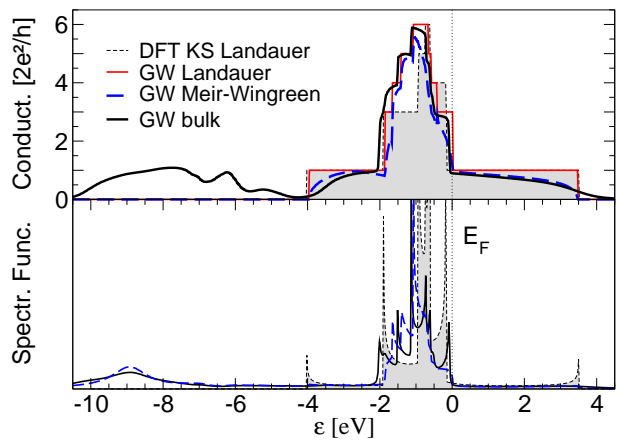}
  \caption{(color online) Conductance (top) and spectral function (bottom) 
  for the stretched atomic configuration. The Fermi level is set to zero.
  Thin dashed line: Landauer result using a DFT KS structure;
  thin solid line: Landauer using only a real part GW renormalization of the energies;
  thick dashed line: Meir-Wingreen result using GW real part renormalization in the leads
  and a full (hermitean+anti-hermitean) dynamical GW self-energy in the conductor;
  thick solid line: GW bulk conductance with full dynamical self-energy.
  }
  \label{cond}
\end{figure} 
In Fig.~\ref{cond} we show the conductance and the spectral function 
of the gold chain for the stretched geometry, obtained using different methods:
The thin dashed line is calculated using the Landauer formula
and the DFT KS electronic structure.
The conductance at the Fermi level for the stretched chain
appears to be one (in units of $2 e^2 / h$) and it is
of $s$-like character. 
This is true also for the relaxed structure (not shown), although
in that case the Fermi level is at the limit of the onset of the $d$-like states.
The thin solid line is obtained from the Landauer formula evaluated using
the GW real-part-only QP energies. GW corrections are considered
both in the conductor and in the leads. Otherwise,
a ficticious contact resistance, unphysical for a homogeneous system,
would appear.
At this first level, the net effect is a renormalization 
of KS into QP energies, the true energies to
introduce and remove an electron from the system.
Therefore, the GWA affects the conductance profile 
by modifying the position of the conductance steps, especially in the
$d$-like region. In the relaxed geometry, the GWA also narrows the $s$-like
conductance channel.

The thick dashed line in Fig.~\ref{cond}
represents the result obtained in the tri-partitioned geometry
by using the Meir-Wingreen formula and
introducing a full non-hermitean and dynamical GW self-energy
in the conductor. Static real-part-only QP energies are included in the
leads.
This introduces loss-of-coherence only in the conductor while leaving
the leads ballistic. At the same time it {\it limits} the introduction of ficticious
contact resistances, {\it i.e.}
the QP levels are aligned in the leads and the conductor.
The difference of this curve wrt the thin solid line genuinely represents
the effect of \textit{e-e} scattering mechanisms in the conductor,
causing diffusion, loss of coherence and appearance of resistance.
With respect to Landauer approaches, the spectral function 
now appears as a collection of broadened QP peaks, whose finite width is
directly associated to the inverse of the electronic lifetime of the QP state.
The spectral weight, which is spread out, results in a lowering
and  a spill-out of the conductance step-like profile.
This effect is directly related to the imaginary part of the QP energies,
and can be seen to increase with $\epsilon - E_{\rm F}$
although not with a quadratical scaling Fermi-liquid behaviour, as 
it is normally observed in GW results for 3D systems.
Finally, we calculate the fully correlated bulk
GW conductance by taking into account 
a non-hermitian and dynamical GW self-energy
everywhere in the system, conductor and leads (thick solid line).
With respect to the previous case, even residual contact resistances
(due to the fact that the conductor and leads spectral peaks were differently shaped,
with finite and infinitesimal widths respectively) are completely removed, and the conductance
increases almost overall.
Only around $-3$ eV we see a slight drop, 
which is due to
the specific $(\Gamma_L + \Gamma_R)^{-1} \Gamma$ factor
in Eq.~(\ref{eq:generalized_landauer}).
Moreover, new structures appear in the conductance at the lowest energies.
By inspecting the spectral function, 
we can attribute them to the presence
of satellites of electronic origin, {\it i.e.} plasmons or shake-ups,
of the main QP peaks.
Since the \textit{e-e} interaction is an elastic scattering mechanism, these satellites 
are necessary to balance the losses which occur at energies close to the Fermi level, 
and are therefore important for transport.
The \textit{e-e} scattering acts in a way to redistribute
the conductance channels to different energies, rather than globally
destroy conductance as in the \textit{e-ph} scattering,
where momentum and current is lost to ionic degrees of freedom.

\begin{figure}
  \includegraphics[clip, width=0.45\textwidth]{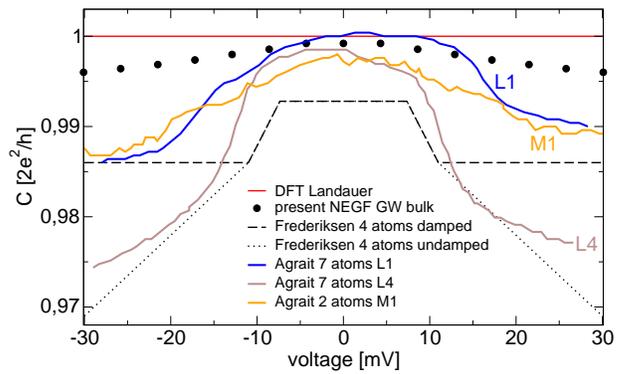}
  \caption{(color online) Differential conductance vs applied bias.
  Thin solid line: DFT Landauer result;
  dots: present NEGF GW bulk result for the 5.35 Bohr interatomic distance;
  dashed and dotted line: \textit{e-ph} theory of 
  Ref.~\cite{FrederiksenBrandbyge2004} corresponding
  to 4 atoms, same interatomic distance and for the damped and undamped limits;
  thick solid lines: experimental result of 
  Ref.~\cite{AgraitYeyati2003} corresponding to 2 and 7 atoms
  and different chain strains.
  }
  \label{characteristic}
\end{figure} 

Taking the GW bulk result, we have integrated the conductance curve
such to obtain the voltage characteristics of
the correlated system. We compare with the experimental results of 
Ref.~\cite{AgraitYeyati2003} and the \textit{e-ph} result of 
Ref.~\cite{FrederiksenBrandbyge2004}, calculated at exactly the same
stretched 5.35 Bohr interatomic distance.
Like in that work, we assume that an
equilibrium picture can still be appropriate to describe
the small voltage range of $\pm$ 30 mV.
The GW result is shown in Fig.~\ref{characteristic} (dots).
The results from Ref.~\cite{FrederiksenBrandbyge2004} 
attribute the step in the conductance,
occurring at $\sim 15$ mV, to the onset of phononic processes.
Instead, the continuous drop observed in our electronic correlated conductance,
occurring in the first 15 mV, 
compares favorably with the drop observed experimentally~\cite{AgraitYeyati2003}: 
\textit{e-e} scattering mechanisms seem hence responsible 
for the conductance drops at very low bias. 
While the quantitative agreement with the experiment on the conductance value
may be somewhat fortuitous~\cite{note-fortune}, 
the trend in this drop is a direct consequence of the
increase in the GW imaginary part of QP energies.

In conclusion, we have calculated the conductance of a realistic gold
chain system by taking into account \mbox{\textit{e-e}} correlation effects
within the GW approximation. With respect to Landauer DFT results,
the conductance profile is considerably modified. 
Already at the level of an equilibrium non-self-consistent GW, 
the trend of the differential conductance appears to compare
favorably with the trend experimentally observed for this system.

We thank L. Reining, D. Feinberg, H. Mera, M. Verstraete, Y. M. Niquet and E. Shirley 
for useful discussions and remarks.
Computer time has been granted by IDRIS proj. 060932.
AF acknowledges funding by Italian MIUR through PRIN 2004.


\begin{thebibliography}{99}
\bibitem{dftdetails} 
Gold $6s$ and $5d$ electrons are kept in valence. 
We adopt a cutoff of 30 Ry for wavefunctions
and a $(8,1,1)$ k-grid.
\bibitem{gwdetails} 
We used 150 and 100 states to converge $W$ and $\Sigma$ respectively;
2503 plane waves have been used to represent the wavefunctions and $\Sigma_x$, while
the dimension of the $\Sigma_c$ and $W$ reciprocal-space matrices
was 49.
Imaginary axis integrations were carried out by a Gauss quadrature using
30 nodes. In the analytical continuation, we used a Pad\'e approximant on
12 nodes. We used a modified version of the \textsc{ABINIT} code.
\bibitem{note-fortune}
In general GW does not achieve such levels of accuracy.
\end{thebibliography}
\end{document}